\documentclass[11pt]{article}

\usepackage[english]{babel}
\usepackage[margin=2.5cm]{geometry}
\usepackage{siunitx}
\usepackage{setspace}
\usepackage{graphicx}
\usepackage{color}
\usepackage{multirow}
\usepackage[affil-it]{authblk}
\usepackage{url}

\usepackage{breakurl}
\usepackage[breaklinks]{hyperref}
\usepackage{nowidow}
\usepackage[sort&compress,numbers]{natbib}
\newcommand{\dropcap}[1]{#1}

\newcommand{\du}{\ensuremath{[\mathrm{1/km^2}]}}

\graphicspath{{./figures/}}

\title{Ride-share matching algorithms generate income inequality}

\author[a,b]{Eszter Bok\'anyi}
\author[c,d,e]{Anik\'o Hann\'ak\footnote{To whom correspondence should be addressed. E-mail: aniko.hannak@wu.ac.at}
} 

\affil[a]{E\"otv\"os Lor\'and University, Budapest, Hungary}
\affil[b]{Agglomeration and Social Networks Lend\"ulet Research Group, Centre for Economic and Regional Studies of the Hungarian Academy of Sciences}
\affil[c]{Complexity Science Hub, Vienna, Austria}
\affil[d]{Vienna University of Economics and Business, Austria}
\affil[e]{Centre for Economic and Regional Studies of the Hungarian Academy of Sciences}

\date{}

\begin{document}

\maketitle

\vspace{-1.5cm}

\begin{abstract}
    Despite the potential of online sharing economy platforms such as Uber, Lyft, or Foodora to democratize the labor market, these services are often accused of fostering unfair working conditions and low wages. 
    These problems have been recognized by researchers and regulators but the size and complexity of these socio-technical systems, combined with the lack of transparency about algorithmic practices, makes it difficult to understand system dynamics and large-scale behavior. This paper combines approaches from complex systems and algorithmic fairness to investigate the effect of algorithm design decisions on wage inequality in ride-hailing markets. We first present a computational model that includes conditions about locations of drivers and passengers, traffic, the layout of the city, and the algorithm that matches requests with drivers. We calibrate the model with parameters derived from empirical data. Our simulations show that small changes in the system parameters can cause large deviations in the income distributions of drivers, leading to a highly unpredictable system which often distributes vastly different incomes to identically performing drivers. As suggested by recent studies about feedback loops in algorithmic systems, these initial income differences can result in enforced and long-term wage gaps. 
\end{abstract}

\dropcap{A}s they grow in popularity, ride-sharing and food-delivery services such as Uber, Lyft, Ola or Foodora are quickly transforming urban transportation eco-systems \cite{Stone2015,Hall2018}. These services have revolutionized most aspects of the transportation market. By managing the rides through a mobile application, they lower the entry barriers to the service for both users or passengers and drivers. The rating system facilitates trust between drivers and users, and the flexible working hours make ride-sharing services a popular choice for people starting a new career or a side-job. 

A key feature of these services is that an algorithm replaces human dispatchers in the task of matching available drivers to the incoming requests. Companies are now able to optimize the matching with unprecedented precision using data they possess on cars, drivers, and traffic conditions \cite{Cheng2011a}, resulting in better service availability, shorter waiting times, and ultimately a boost in company profits \cite{Kedmey2014}. On the other hand, in the process of maximizing profit, drivers' interests get sidelined, and undesirable social outcomes can emerge \cite{Popper2014, Shepardson2019, Oh2014,Griswold2014}.

Recent studies raise concerns about the risks threatening workers’ well-being, including racial bias, worker safety, fairness to workers, and asymmetries of information and power \cite{fairwork,Graham2017,Graham2018, Lee2015a, Cook2018}. Since workers are not able to obtain remedies through official channels \cite{confessions_tumblr, Rosenblat2016}, strikes have gotten increasingly common in the past years. Drivers of Uber, Lyft, Ola, Foodora demand higher fares, job security, and livable incomes all over the world \cite{Dickey2019}. 



While traditional taxi services allowed all workers to hear the same information and talk to the dispatcher, in these modern systems drivers see nothing besides their next target \cite{Rosenblat2016}. The automatized algorithmic features act as a barrier between employees and management, annihilating this relationship and ultimately preventing worker feedback to be incorporated into the design. Moreover, the proprietary nature of these systems, there is limited access to data or the rules of the system \cite{Popper2014, Clark2015, Wilson2015}. Both regulators and researchers struggle to gain a deeper understanding of the ongoing problems and the underlying processes, which ultimately hinders informed intervention or monitoring by regulatory agencies \cite{Chen2015, Jiang2018}. 

Most existing literature in the area of taxi matching algorithms is concerned with optimizing aggregate outcomes for the whole system \cite{Kummel2016, Jorgensen2007,Alshamsi2009, Bailey1987, Maciejewski2015, Maciejewski2015b, Maciejewski2016}. This approach aims to maximize the benefits for the company or to minimize the adverse effects such as $\mathrm{CO_2}$ emissions, overall distances driven or the passenger waiting times. Following the line of fairness measurement literature \cite{Hutchinson2018, Chakraborty2017, Zehlike2017, Weng2017}, we instead focus on the fair distribution of income from the drivers’ perspective, because current systems do not guarantee the same income for the same amount of work, neither across workers nor over time \cite{Lee2015a,Fieseler2017,Graham2017,Graham2018}. 

 

We use an agent-based simulation to systematically study the mechanisms in ride-hailing and delivery systems from the perspective of the drivers. We first quantify the income inequality level of systems that use a company-level profit maximizing approach, and explore how the system-level behavior changes as a function of input parameters like the number and distribution of taxis and passengers, city layout, and driving strategies. Next, we investigate the trade-off between fairness of driver incomes and loss in overall revenue through an algorithm, designed to integrate the fairness perspective into the matching of drivers with requests.




\section{Income Inequality in Ride Hailing}

We create a simulated city environment with drivers, passengers and various parameters which we will introduce throughout the section. Our goal is to simulate a diverse range of cities and real-world traffic conditions, thus we derive our model parameters from real-world data sets.

Taxis in our city drive along a grid, moving one block at each time step. In our basic setting both request and drop-off locations are most likely to be in the city center, in line with previous studies on real-world data \cite{Qian2015, Peng2012, Liu2012}. The matching algorithm between drivers and requests is similar to the algorithm that Uber and most taxi companies use: passengers are matched with the closest available car. Additional parameters capture various real-world scenarios such as city layout, changes in supply and demand, driver strategies, and different settings of the matching algorithm. The pricing scheme is similar to that of UberX in Boston \cite{uberboston}, and fuel costs are accounted for \cite{gasprice}. We run all simulations for what equals a 40-hour work week (see \ref{sec:mm} for details).

As described before, major complaints of drivers are low wages, and the unpredictability and unfair distribution of incomes. Low wages could only be addressed by the company itself, because they are in control of the payment system. Thus, we focus on understanding the two other complaints through investigating the distribution of incomes across workers for various settings of our system.

\subsection{Changes in Supply and Demand}

We use two variables to calibrate the supply and the demand in our system: 
the supply $d$ (density) is defined as the number of taxis per square kilometer, while $R$ captures demand-to-supply ratio, which is the fraction of the demanded travel distance over the supplied travel distance (see \ref{sec:mm} for details). 
To pick parameter spaces for $d$ and $R$ that cover actual real-world scenarios, we calculate the number of cars and passengers using empirical data from the NYC Taxi and Limousine Commission \cite{nyctaxitrips}.

\begin{figure*}[t]
    \centering
    \includegraphics[width=\linewidth]{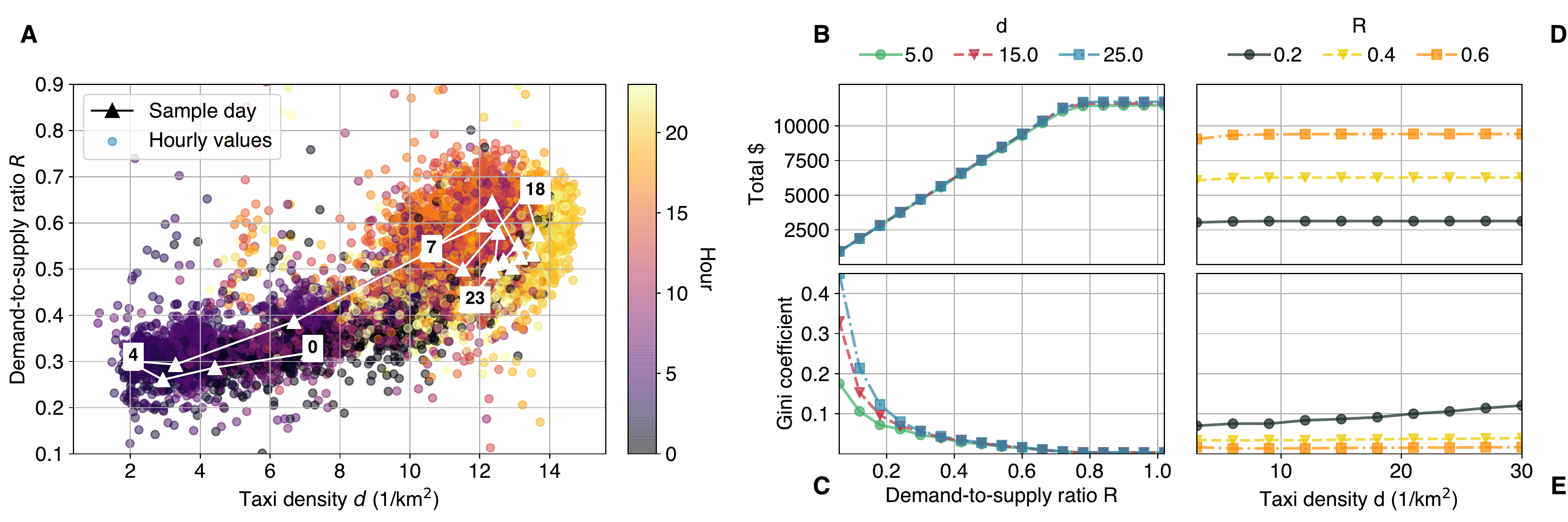}
    \caption{(A) Hourly values of the taxi density $d$ and the demand-to-supply ratio $R$ throughout year 2013 in NYC. Parameter changes within one selected day, 15 January are plotted as white connected triangles. The total income (B) and Gini coefficient (C) for varying demand to supply ratio, at three fixed density values. We see that income grows linearly as the demand to supply ratio grows. Higher demand leads to decrease in inequality, for all density levels. The total income (D) and Gini coefficient (E) for varying density values, at three fixed demand to supply ratio values. Higher demand generates higher incomes. Inequality grows with more cars in a city. }
    \label{fig:fig1}
\end{figure*}

Figure~\ref{fig:fig1}A depicts each hour of year 2013 in NYC in the $d-R$ parameter space. Plotting an example day, 15 January 2013, from the same dataset with its hourly parameter values shows, that even within one day and one city, both parameters strongly vary. While such detailed data is not available from other cities, we analyzed aggregate statistics as well as extreme events such as public transportation failures, strikes or bad weather conditions (see Section~\ref{sec:mm}). We conclude that the taxi density range $d\in\left[3,\, 30\right]$~\du~and the demand-to-supply ratio range $R\in\left[0.1,\, 1\right]$ will capture a wide range of city sizes, seasonal changes, varying traffic at different times of the day, and even extraordinary events that cause sudden changes in the demand or supply.

Figure~\ref{fig:fig1}B shows how the average income changes with growing number of requests, but constant taxi numbers, that is, a constant density $d$. Each marker denotes the averaged result of 10 simulations, ran with the associated parameters. As expected, the income is directly proportional to the demand, regardless of $d$. For example by doubling the demand-to-supply ratio $R$ from 0.3 to 0.6, the income roughly increases from \$4700 to \$9400 for all three $d$ values. Thus, as long as the system is able to serve all the passengers, demand determines the total income. After a certain point, incomes saturate as taxis are not able to serve all requests, and the system reaches its maximum capacity around $R=0.78$. \footnote{
We caution the reader that fares highly vary across cities, service providers, and over time even within the same company. While we present the incomes in \$, the values are approximations of the fares and the emphasis is on the relative value and the shape of the distributions. 
}

Figure~\ref{fig:fig1}C measures the inequality for the same parameters using the Gini coefficient of the incomes at the end of the simulation. Gini coefficient is an inequality measure that captures the deviation of the Lorenz curve of the income distribution from that of an ideal one, where a given cumulative percentile of the population holds the same percentile amount from the incomes (see Section~\ref{sec:mm}). For low demand, that is, low $R$, the Gini coefficient starts at high values, with $0.18$ for $d=5~\du$, $0.32$ for $d=15~\du$ and $0.45$ for $d=25~\du$. As the demand increases, the Gini coefficient decreases and converges for different taxi densities, with the Gini at $R=0.6$ equalling to as little as 0.01 for all three measured $d$ values. 

Figures~\ref{fig:fig1}D-E depicts the effect of increasing traffic (more demand and more supply), that is, constant $R$, but increasing taxi density $d$. Again, higher demand generates higher average incomes, that is roughly \$3100 for $R=0.2$, \$6200 for $R=0.4$, and \$9400 for $R=0.6$. The income is only very slightly affected by the taxi density in a city, as it increases by only 3\% for a tenfold increase of $d=3~\du$ to $d=30~\du$ . However, higher density leads to a more unequal distribution of those same average incomes, as seen from Figure~\ref{fig:fig1}E. For $R=0.2$, the Gini coefficient almost doubles with the same a tenfold increase of taxi density $d=3~\du$ to $d=30~\du$ that caused the 3\% increase in the incomes. For higher demand, this inequality shift is smaller: while there is a 14\% increase with  changing $d=3~\du$ to $d=30~\du$ for $R=0.4$, the Gini is almost constant for $R=0.6$.

We note, that the Gini coefficients for the high supply (high $d$) and low demand-to-supply (low $R$) regions are comparable to the Gini coefficients observed for country-wise incomes \cite{Gasparini2015}. Given that national statistics encompass inequalities across different professions and sectors, and we only compare workers of the same sector, with the same working hours and even with the same fitness, these values translate to a surprising level of inequality. 


\subsection{Varying Spatial Activity}

\begin{figure*}[ht]
    \centering
    \includegraphics[width=\linewidth]{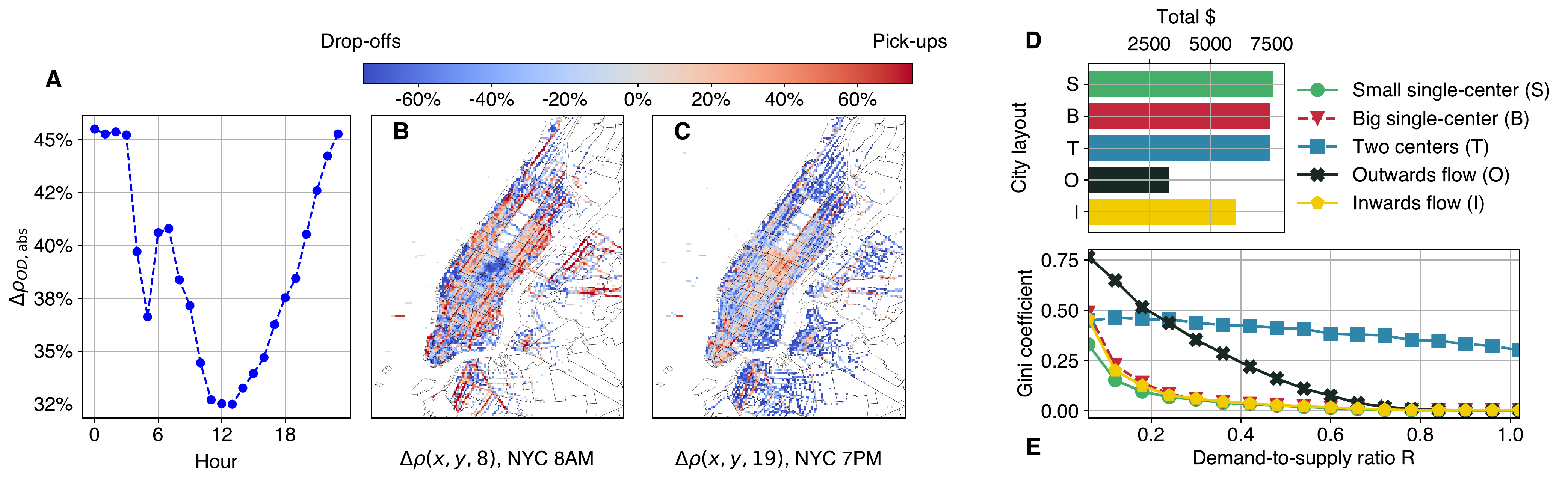}
    \caption{(A-C) Temporal and spatial changes in the difference between trip origin and destination locations in NYC. (A) Hourly average of the unsigned symmetric difference $\Delta \rho_{OD,\mathrm{abs}}$ in NYC pixel areas. The figure indicates that origin and destination patterns vary throughout one day. (B)-(C) Spatial distribution of $\rho_{OD}$ at 8AM and 7PM in NYC. Red values indicate more pickups, blue values more drop-offs in a pixel area. At 8AM, there is an influx into the center of Manhattan and other business and work areas. At 7PM, the direction of the taxi flow is reversed, as people travel towards the outskirts from the center. See Section~\ref{sec:mm} for details on $\Delta \rho_{OD,\mathrm{abs}}$ and $\rho_{OD}$. (D) Average income of drivers depending on city layout. Generally, the overlapping cases generate more income, and movement towards the city center as well. (E) Gini coefficient of the seven layouts with varying demand to supply ratios. As demand grows, inequality decreases for every layout. Non-overlapping layouts (flows) in the city generate higher inequality.}
    \label{fig:fig2}
\end{figure*}

Next, we consider variations in the city layout and the traffic flow patterns. Apart from the simplest city-center scenario, we run simulations for i.) a city with a larger center, ii.) a city with 2 centers, and iii.) cases where the pickup and destination locations do not to overlap (see the detailed description of these layouts in Section~\ref{sec:mm}). Our motivation is explained by Figure~\ref{fig:fig2}A which shows the temporal and spatial changes in the difference between trip origin and destination location distributions in NYC. We can clearly see the average flow of passengers in the morning towards the center for work (Figure~\ref{fig:fig2}B) and away from the center in the evening  (Figure~\ref{fig:fig2}C).

Before we turn our attention to income inequality in the above layouts, it is worth investigating whether and how they affect the overall income, since they all share the same $R$ given identical traffic, i.e., the same demand-to-supply ratio. Figure~\ref{fig:fig2}D shows that the average income strongly depends on the distribution of spatial activity, when examining a setting with fixed $d=15~\du$ and $R=0.48$. If the distribution of origins and destinations overlap completely, more income is generated regardless of the details, e.g. the spreading of the city center or having multiple centers. On the other hand, in the cases where there is a dominating flow of vehicles due to asymmetries in the origin and destination demands, the income is lowered to roughly 50\% (Outwards flow) and 20\% (Inwards flow). Moreover, the inequality is significantly higher over the whole demand-to-supply $R$ range (Figure~\ref{fig:fig2}E). In the Inwards flow case, the Gini coefficient starts from as high as 0.76, but decreases fast thereafter, and reaches 0.07 for $R=0.6$. When the traffic flows mostly outwards from the city-center, the initial Gini of 0.45 decreases very slowly, though, and stays as high as 0.38 for $R=0.6$. The above results indicate that the distribution of spatial activity strongly influence both the average income and the income inequality.

\subsection{Waiting vs Cruising}
An important decision every driver faces many times throughout the day is what to do in the time period until the next request comes in, which might amount to a considerable share from the overall time spent online \cite{Cheng2011a, Jiang2018}. So far in our simulations, the cars were waiting in place after dropping off their passengers. Another obvious idling strategy is to cruise back towards the center to meet more demand. While in a real system, drivers presumably use a mix of these two strategies, here, we investigate the outcome of the two extreme cases, namely when all taxi drivers either wait or head towards the city center.

Intuitively, we would expect that the cruising strategy leads to similar throughput in the cases of overlapping pick-up and drop-off distributions, smaller overall throughput in the cases where traffic flows towards the center, and higher overall throughput when population flows towards the outskirts. But does higher throughput also lead to lower Gini coefficients?

\begin{figure}[h]
    \centering
    \includegraphics[width=8.7cm]{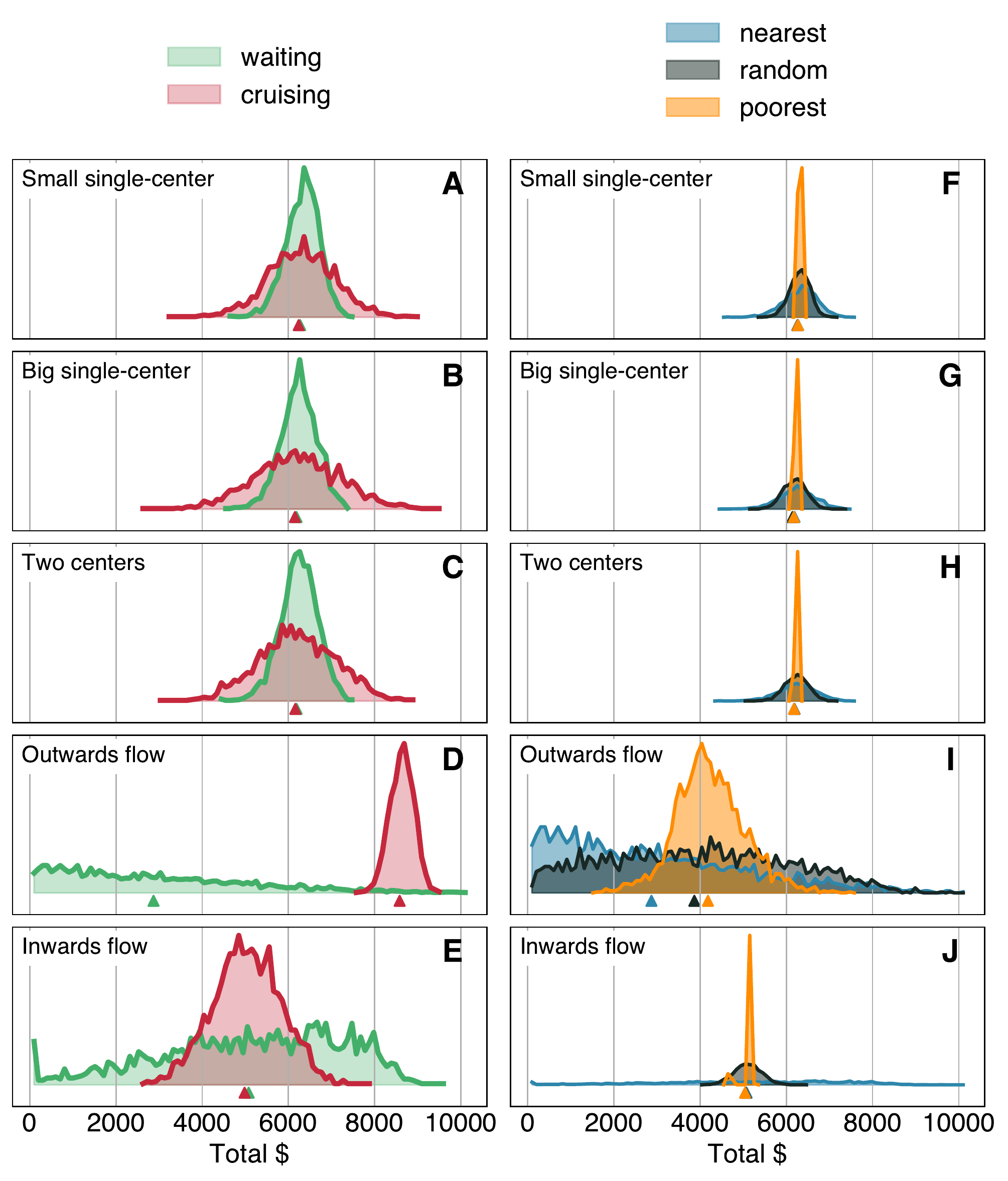}
    \caption{(A-E) Distribution of incomes of different cruising strategies for different city layouts at $R=0.4$ and $d=~15\du$. In symmetrical layouts the cruising strategy leads to more unequal distribution of incomes with similar means. Among the asymmetrical layouts, in the case of outwards flows, cruising towards the center leads to higher incomes. (F-J) Distribution of incomes of different matching algorithms in different city layouts and with the waiting strategy, at fixed $d=15~\du$ and $R=0.4$. The poorest matching strategy creates a more equal distribution for all of the investigated geometries. (A-J) The vertical scale of the distributions is omitted for better readability, but all distributions are normalized such that the areas under the curve are equal to 1. Triangles mark the means below the distribution curves.}
    \label{fig:fig3}
\end{figure}

Figure~\ref{fig:fig3}A-E shows the waiting/cruising scenarios side by side for $d=15~\du$ and $R=0.4$. We observe significant differences in fairness between the waiting and cruising strategies in all city layouts. In the case of overlapping pick-up/drop-off location distributions the waiting strategy is fairer, as illustrated by the narrower distributions on Figure~\ref{fig:fig3}A-C corresponding to lower Gini coefficients while the average income is untouched. In case of the asymmetrical layouts, the strategy of cruising back to the center increases fairness, and in the Outwards flow layout it even raises the average income (decrease of the Gini from 0.22 to 0.07 and average income increases by almost 200\%). 

These results underline the importance of transparency and the direct effect of information asymmetry on drivers, who in the current setup of ride-hailing systems can not make informed decisions about their strategies. Moreover, it shows that a seemingly small changes in the system settings can lead to large differences in the fairness guarantees of the overall system. 

\subsection{Fixing the algorithm}

Lastly, we examine whether we can incorporate the fairness perspective into our system and achieve more equal incomes while conserving the overall revenue, similarly to the perspective of \cite{Chakraborty2017}. Our goal is to keep track of drivers' income throughout the day and take it into account when assigning rides.

With this idea in mind, we create the \textit{poorest} algorithm: a modification of the current matching algorithm which keeps track of drivers' income at each point in time and assigns taxis not only based on distance from the potential passenger but also the money they made so far (see details in Section~\ref{sec:mm}). To meaningfully compare the algorithms, we include a baseline algorithm which assigns drivers to passengers randomly within a given radius. This \textit{random} setup should create higher fairness but lower total income than picking the nearest available vehicle. 

Figure~\ref{fig:fig3}F-J shows the income distribution for the three algorithms for all city layouts. The narrow distributions of the \emph{poorest algorithm} in the symmetric cases show that this poorest correction effectively mitigates the adverse effects of the \emph{nearest algorithm}. In these cases, \emph{poorest} performs even better than the \emph{random assignment} that we consider as a baseline for fair conditions. Moreover, the strategy also helps mitigating inequalities on the Inwards flow layout. While not as strongly as with the other layouts, poorest significantly increases fairness with the Outwards flow layout, and even increases the mean income. Since the nearest algorithm mostly assigns drivers from the center, more and more drivers end up in the outskirts without close-by rides. As the poorest algorithm is more likely to pick a driver stranded on the outskirts, it compensates this undesired process and ultimately leads to higher mean income. 

\section{Concluding Discussion}

We presented results of an agent-based automated taxi system simulation from a so-far unexplored angle, namely, the fairness of incomes among drivers. Our simulation environment allows us to cover a wide spectrum of city layouts and traffic conditions, and to test arbitrary matching algorithms. We find that a low demand-to-supply ratio (more taxis/less passenger requests) leads to greater inequality, and that the inequality is largely dependent on the spatial distribution of request origins and destinations, on the chosen idling strategy of the drivers, and the matching algorithm. 

We also proposed a new matching algorithm that attempts to equalize incomes by promoting drivers on the lower end of the current income distribution. This method significantly boosts fairness in driver incomes in all investigated scenarios, while not impairing, and sometimes even increasing average incomes through stimulating optimal spatial redistribution of the taxis.

Notably, our most surprising result is that drivers with the same qualifications and working hours can end up with vastly different incomes by chance. One might argue that these income differences, observed over a short time period, disappear on the long run. This would indeed be the case under our model if used for long-term predictions (even if the regression to the mean could take a considerable amount of time). 
However, a large body of sociological and economic literature has shown that emerging inequalities are amplified through feedback loops and other processes that are not directly captured by the simulation in this paper \cite{Cook2018,van2014field}. The initial differences combined with emerging feedback loops and varying skills, working hours, and ratings can result in increased and continuing wage gaps.
Moreover, literature shows that workers have to focus on daily and weekly targets because of regular fees and payments \cite{Farber2005}. Thus, short-term income variance creates unpredictable circumstances and may cause drivers to take risks such as unsafe amounts of overtime \cite{Rosenblat2016}.



Due to the sheer size and complexity of the problem, our study is limited in different ways. First, the city model generalizes to a square grid street network, and the traffic model averages over highly dynamic conditions such as the true taxi velocity, which might depend on the time of day, road type, and physical obstacles. For example, Uber's surge pricing, which aims to steer supply and demand, is an important factor in driver earnings and another source of unpredictability. In contrast, our algorithms are only elementary dispatching systems that are likely not identical to the proprietary solutions.

\paragraph{Impact}


Our results underline the urgency to establish the necessary means for drivers to make informed decisions while working. In our view, giving more control to workers is only helpful when coupled with insights that help them make informed decisions. Currently, drivers' basic decisions, such as when and where to work, are based on limited and local information. Our systematic analysis aims to fill this gap by connecting local knowledge with global outcomes while focusing on drivers' interests. We derive results that directly help drivers align short-term strategies with their long-term goals based on a few key system indicators such as demand, supply or the spatial distribution of requests.



Our work will be the first to take fairness metrics into account in the evaluation of system performance. Currently, companies focus on maximizing their own profit, which does not align with worker welfare and fairness. Technical research has followed the former demand, considering only optimization problems of system level efficiency. By sharing our simulation system on GitHub\footnote{\url{https://github.com/bokae/taxi}}, we hope to encourage the research community to work towards a thorough understanding of the effects of algorithm choice on social outcomes, and that eventually service designers will be provided with specific technical recommendations on how to implement a possible fairness perspective. Ultimately, fairness guarantees, either provided by the platforms themselves or via external monitoring of policy agencies, are necessary to increase the trust between ride-hailing companies and their workers. 

Lastly, we want to emphasize the advantages of complex systems methods in algorithmic fairness research. Currently, most algorithmic auditing research projects focus on identifying issues with individual platforms at particular points in time \cite{sandvig2014auditing,hannak2013measuring}. We presented an enhanced approach to algorithmic auditing that takes into account system dynamics, complex phenomena, and downstream consequences of algorithms. This allows the generalization of insights across platforms, pointing towards fundamental properties of socio-technical systems. Complex algorithmic auditing has the potential to be adapted to research harmful phenomena beyond this project, such as the filter bubble effect, discrimination, or polarization in algorithmically aided systems \cite{flaxman2016filter,hannak-2017-cscw,conover2011political}.

\section{Materials and Methods}
\label{sec:mm}

\paragraph{Simulated city} We investigate generalized and simplified ride-sharing systems with an agent-based simulation. In this simulation, the city is represented as a square grid similarly to \cite{Ranjit2018}, consisting of $n$ times ($n=20$) $m$ pixels ($m=20$). To associate simulation units to real-world dimensions, the side length of the pixels is set to $du=\SI{100}{m}$ in SI distance units. 

\paragraph{Time, movement and distances} In each simulation time step, taxis move exactly one distance unit on the grid, and they keep this constant average velocity $v$ throughout the whole simulation. We set $v$ to be $\SI{10}{m/s}$, a hypothetical average daytime speed as in \cite{Kummel2016}, which implicitly defines simulation time unit: $tu = a/v = \SI{10}{s}$. All presented results throughout the paper correspond to simulations ran for what equals a 40-hour work week, that is, until $5\cdot 8\cdot 3600 /tu = 14 400$ simulation time units. Literature suggests one work week to be a reasonable time frame for which workers might expect predictable and stable incomes because of weekly fixed costs \cite{Farber2005}. 


\paragraph{Supply and demand} There are $T$ taxis in the system. We control the number of taxis through a density parameter $d$, that is the number of taxis per total area of the system. The dimension of $d$ is \du.
\begin{equation}
    d = \frac{T}{A} = \frac{T}{n\cdot m\cdot du^2}.
    \label{eq:d}
\end{equation}

The demand-to-supply ratio is characterized by a dimensionless measure $R$, which is the fraction of the total demanded travel distance over the distance provided by the taxis during the same time window:
\begin{equation}
    R = \frac{\lambda\cdot l \cdot t}{T\cdot v \cdot t} = \frac{\lambda \cdot l}{T\cdot v},
    \label{eq:l}
\end{equation}
\noindent where $l$ is the average trip length requested by passengers in the system, $t$ is the elapsed time, and $\lambda$ is the average number of requests per time unit. 

\paragraph{Traffic flow and city layout} Since the spatial distribution of the demand is typically not uniform in a city \cite{Qian2015, Peng2012, Liu2012}, we introduce different probability distributions that characterize the request origins and destinations. In the simplest case, the origin and destination distributions are identical 2D Gaussians, mimicking scenarios where demand is higher in the city center, and gradually decreases further from the core \cite{Yuan2011, Liu2012b}. The scenarios where origin and destination distributions are not overlapping, such as in the case of morning rush hours or late night rides to residential areas, are simulated by defining separate Gaussian distributions for request origin and destination, respectively, with different standard deviations or centers. Table~\ref{tab:geometry} provides the different distributions in our simulations, with $f$ being the weight of Gaussian mixtures, $\mu$ their centers, and $\sigma$ their standard deviations. We compared the distribution of trip lengths generated from these Gaussian distributions to that of a real-world dataset \cite{Wong2014}. We concluded that the distribution from a real day is the composition of many different system states and spatial activities, therefore, we do not require a strict match between the statistics of the two. We also generated layouts with different cutoffs (exponential, sigmoid) in the decrease of the density of the pick-ups and drop-offs towards the outskirts, but the results did not differ qualitatively from those obtained with Gaussian distributions.

\def\arraystretch{1.5}
\begin{table}[h]
    \centering
    \begin{tabular}{l|ccc|ccc}
        & \multicolumn{3}{c|}{Origin} &  \multicolumn{3}{c}{Destination} \\
        & $\mu$ & $\sigma$ & $f$ & $\mu$ & $\sigma$ & $f$ \\ \hline \hline
        Small city-center & [20,20] & 10 & 1.0 & \multicolumn{3}{c}{$-$} \\ \hline
        Large city-center & [20,20] & 20 & 1.0 & \multicolumn{3}{c}{$-$} \\ \hline
        \multirow{2}{*}{Multi-centered city} & [12,12] & 8 & 0.5 & \multicolumn{3}{c}{\multirow{2}{*}{$-$}}  \\
            &[28,28] & 8 & 0.5 \\ \hline
        Outwards flow &  [20,20] & 6 & 1.0 &  [20,20] & 12 & 1.0 \\ \hline
        Inwards flow & [20,20] & 12 & 1.0 &  [20,20] & 6 & 1.0 
    \end{tabular}
    \caption{List of different origin and destination distributions in the simulations. Each three-column row represents a 2D Gaussian distribution with a mean at $\mu$ in simulation coordinates, a standard deviation $\sigma$, and a weight $f$. Distributions having more than one row are a mixture of the listed Gaussians with a weight $f$. The $-$ symbol denotes that the destination distribution is the identical to the origin distribution.}
    \label{tab:geometry}
\end{table}

\paragraph{Implementation} Once we have the taxis and incoming requests in the system, we have to define how these elements behave while running the simulation. If a passenger makes a request, the origin and destination location is submitted to the waiting queue, and the request is first marked as pending. At each time step, we try to assign an available taxi to all requests in the queue, from oldest to newest. If the assignment is successful, the request is marked as confirmed, and the assigned taxi is marked as booked, and it ceases to be available. When the taxi reaches the request origin, it picks up the passenger, and the taxi and the request both enter into serving state. A path is created for the taxi, along which it moves towards the request destination as the time is running. Once the destination is reached, the taxi drops off the passenger, the request is marked as completed, and the taxi becomes available again. Requests that pend too long, that is, for more than 30 simulation units (corresponding to \SI{5}{min} in real time units) in the pending queue are canceled, as passengers tend to not waiting very long.

\paragraph{Income} Because the pricing scheme might differ between different service providers, cities, or it might change rapidly even within the same system (similar to surge pricing in Uber \cite{Diakopoulos2015}), we only introduce illustrative USD values based on real UberX fares from an example US city, Boston \cite{uberboston}. Thus, the income has a fixed \$2/trip and a distance-based \$1/km part, that already includes the mile-based and minute-based fees calculated using an average velocity of \SI{10}{m/s}. Note that the final income values only serve as indicators of general income trends in mean incomes and inequalities, and they shall not be interpreted as realistic incomes. Fixed costs are assumed to be identical for each driver, thus they are omitted from the calculations. Fuel consumption is proportional to the total distance covered (going for and serving a request, cruising while empty). With an average  price of fuel of \$3 per gallon \cite{gasprice}, an estimated \SI{10}{l} per \SI{100}{km} \cite{Weng2017} consumption leads to $\approx\$0.008/du$ costs.

\paragraph{Matching algorithm} The essence of the assignments between taxis and requests is the matching algorithm. All presented algorithms first consider a pool of available drivers in a \SI{1}{km} range around the request origin. First, we implement the \emph{nearest} algorithm, that looks for the nearest available taxi from the pool. This algorithm is the basis for many existing solutions \cite{Novak2013, Maciejewski2015, Maciejewski2016, RobinsonLaRocca2017}. We also propose a mechanism that takes the drivers' previous incomes into account and the driver with the least income so far (\emph{poorest}). As a baseline, we also include a method with uniform random assignment from the pool (\emph{random}).

We can also set different initial conditions and different idle behaviour for the taxi drivers. They either start the day from a "taxi base" in the center location, or from a random location on the grid that mimics their "home" location. After a completed ride, they can either be \emph{waiting} for their next ride in place, or they can start \emph{cruising} back towards the city center. There is also a possibility for simulating a realistic 8-hour shift, where after the end of the shift, every taxi completes its last assignment (if it has one), and then starts the new shift again from their "home/taxi base".

The code for the simulation is available at \url{https://github.com/bokae/taxi}.

\paragraph{Supply and demand parameter ranges}

To match our simulation to real-world parameter values, we obtained the only comprehensive ride-sharing vehicle dataset that is available for the vehicles operating in New York City from year 2013 \cite{Wong2014}. We calculate average trip length $l$, average velocity $v$, number of distinct taxis operating $T$, and number of served requests per second $\lambda$ for each hour of the year from the dataset. Using equations (\ref{eq:d}) and (\ref{eq:l}), we calculate the two main system parameters $d$ and $R$ for these hours using $A=\SI{783}{km^2}$ as the area for NYC. Figure~\ref{fig:fig1}A shows the calculated values, and the changes in $d$ and $R$ within one sample day on 15 January 2013. According to the figure, $d$ varies between 2 and 15~\du, and $R$ between 0.2 and 0.75. However, we do not have data on the number of cancelled requests. Therefore, it might happen even within this dataset, that the real value of $R$ is higher than 0.75 during peak hours. We also calculate approximations of $d$ and $R$ based on aggregated metrics for other towns. For example, we take a typical example for an oversupplied taxi system, the traditional taxi service in Berlin, with $d\approx 1-5~\du$ throughout one day, and a typical $R=0.5$ in peak hours \cite{Bischoff2015}. An opposite scenario of undersupply is present in the city of Barcelona, where $d$ varies between 15~\du and 35~\du, and $R$ between 0.15 and 0.5 \cite{Maciejewski2016b}. Though representing very different real-life situations, in terms of overall numbers, both cities fall into our simulated parameter range. Extreme events such as public transportation failures or bad weather conditions might cause a sudden increase in demand, sometimes as high as doubling $R$ \cite{Maciejewski2016b}. Therefore, we conduct our simulations for up to $R=1$.

\paragraph{City layouts}
We also investigate whether an expected shift in the direction of the trips happens during a day. First, we divide NYC into rectangular pixels of 0.01$^\circ$ longitude and 0.01$^\circ$ latitude, and we calculate the normalized distribution of pick-ups $\rho_O(x,y,h)$ and drop-offs $\rho_D(x,y,h)$ for each hour $h=0,\dots,23$ averaged over one year. We filter those $x,y$ longitude-latitude pixels in the dataset, where the overall number of both pick-ups and drop-offs is below 5, and where $x\notin[-74.05,-73.9]$, and $y\notin[40.67,40.85]$.

Normalization means that for each $h$,
\begin{equation}
    \sum_{x,y}\rho_O(x,y,h) = \sum_{x,y} \rho_D(x,y,h) = 1.
\end{equation}

We then define the symmetric difference between these two distributions:
\begin{equation}
    \Delta \rho_{OD}(x,y,h) = \frac{\rho_O(x,y,h)-\rho_D(x,y,h)}{\rho_O(x,y,h)+\rho_D(x,y,h)}
\end{equation}

Figure~\ref{fig:fig2}A shows the daily mean of the absolute values of the symmetric differences in the pixels
\begin{equation}
   \Delta \rho_{OD,\mathrm{abs}}(h) =  \frac{1}{N}\sum_{x,y} |\Delta \rho_{OD}(x,y,h)|,
\end{equation}
where $N$ is the number of $x,y$ pairs in the analysis. Figure~\ref{fig:fig2}B and C show $\Delta \rho_{OD}(x,y,8)$ and $\Delta \rho_{OD}(x,y,19)$, respectively, on the taxi zone map of NYC. We can observe the inflow of people into the city center in the morning, and the outflow of the traffic around 7PM.

\paragraph{Fairness metric and its interpretation}

To assess the inequality in the income distributions of the drivers after 1 week's worth of work, we calculate the Gini coefficient \cite{Champernowne1974}. The Gini coefficient is an important tool in economics for analyzing income distributions over countries or regions. Its typical value for countries (being between 0.2-0.4) is summarized e.g. in OECD reports \cite{Gasparini2015}. The value of the Gini coefficient is independent of the mean of the distribution, therefore, differences in the inequalities can be compared for scenarios generating different mean incomes in the simulation.

\section*{Acknowledgements}

We would like to thank the helpful comments and insights of Johannes Wachs, Christoph Stadtfeld, S\'andor Juh\'asz, G\'abor Hann\'ak, Kenneth Joseph, Piotr Sapiezynski and David Garcia. During this research, Eszter Bok\'anyi was supported by the \'UNKP-18-3 New National Excellence Program of the Hungarian Ministry of Human Capacities. Aniko Hannak acknowledges funding from the Russell Sage Foundation (92-17-03).

\singlespacing
\bibliography{pnas-sample}

\end{document}